# NMR Spectra Denoising with Vandermonde Constraints

Di Guo, Runmin Xu, Jinyu Wu, Meijin Lin, Xiaofeng Du and Xiaobo Qu[*]

*Abstract*—Nuclear magnetic resonance (NMR) spectroscopy serves as an important tool to analyze chemicals and proteins in bioengineering. However, NMR signals are easily contaminated by noise during the data acquisition, which can affect subsequent quantitative analysis. Therefore, denoising NMR signals has been a long-time concern. In this work, we propose an optimization model-based iterative denoising method, CHORD-V, by treating the time-domain NMR signal as damped exponentials and maintaining the exponential signal form with a Vandermonde factorization. Results on both synthetic and realistic NMR data show that CHORD-V has a superior denoising performance over typical Cadzow and rQRd methods, and the state-of-the-art CHORD method. CHORD-V restores low-intensity spectral peaks more accurately, especially when the noise is relatively high.

## I. INTRODUCTION

NUCLEAR magnetic resonance (NMR) spectroscopy is an important technique that can determine the molecular structure and physical properties of substances in chemistry, biology and medicine [1]. The time domain signal of NMR spectra are called the Free Induction Decay (FID) signal [2], which are often corrupted by noise in practice. To improve signal-to-noise ratio (SNR), a common approach is to average signals obtained in multiple acquisitions but also suffers from longer data acquisition time. Thus, denoising is important to suppress noise.

NMR spectra denoising aims to restore a clear signal from the noisy FID. Typical methods include wavelet denoising [3], linear filters [4], Cadzow signal enhancement [5-8], truncated Singular Value Decomposition (SVD) [9][10], randomized QR decomposition (rQRd) [7], and the Convex Hankel lOw-Rank matrix approximation for Denoising exponential signals (CHORD) [10]. The last four methods share the same assumption that the FID can be modeled as the sum of damped exponential signals [2, 11-13].

This exponential signal assumption is classic and effective in NMR data processing [2, 11-13]. A common process is to convert the FID into a Hankel matrix, whose matrix rank is equal to the number of exponentials in FID. If the number of exponentials (or spectral peaks) is smaller than the matrix, the Hankel matrix will be low rank. This nice property implies that the rank does not depends on the signal intensity and have been evidenced powerful in recovering challenging low-intensity signals [2, 12, 14-16].

For both Cadzow and truncated SVD, the Hankel matrix is decomposed to obtain singular values, and this processing is time-consuming when the Hankel matrix size is large[7][14][17]. QR decomposition turns to a randomized matrix with smaller size but the same rank of the original Hankel matrix, thus significantly decreases the computation time. For both Cadzow and rQRd, it is necessary to estimate the number of exponentials, i.e., the number of spectral peaks. Thus, their performances are limited by the accuracy of estimation. If the parameter is selected too large or too small, noise residue or peak loss may occur. By formulating the denoising as low-rank regularization problem, Qiu *et al*. [10] recently proposed an automatic denoising method called CHORD. It has obtained more robust and accurate results than Cadzow and rQRd, but may produce some pseudo-peaks and residual noise at high noise levels.

In this paper, we follow the same signal assumption as CHORD, and add another matrix decomposition constraint, the Vandermonde decomposition [15, 18]. The Vandermonde decomposition fits well for exponential functions since each exponential will be explicitly decomposed into a single signal subspace. The denoising is formulated as a minimization problem of the norm of exponential functions and the low-rankness of Hankel matrix. Comprehensive results on synthetic and real NMR spectroscopy data are used to demonstrate the advantages.

The rest of this paper is organized as follows. Related works are introduced in Section II, the proposed signal reconstruction model and its numerical algorithm are described in Section III. Then, experimental results and conclusions are provided in Sections IV and V, respectively.

This work was partially supported by the National Natural Science Foundation of China (62371410, 62122064, 61971361 and 62331021), the Natural Science Foundation of Fujian Province of China (2021J011184 and 2023J011427), the President Fund of Xiamen University (20720220063), and Nanqiang Outstanding Talent Program of Xiamen University. (*Corresponding author: Xiaobo Qu, Email: quxiaobo@xmu.edu.cn)

D. Guo, R. Xu and X. Du are with School of Computer and Information Engineering, Fujian Engineering Research Center for Medical Data Mining and Application, Xiamen University of Technology, Xiamen 361024, China.

J. Wu and X. Qu are with Department of Electronic Science, Fujian Provincial Key Laboratory of Plasma and Magnetic Resonance, Xiamen University, Xiamen 361102, China.

M. Lin is with Department of Applied Marine Physics and Engineering, Xiamen University, Xiamen 361102, China.

## II. Preliminaries and Related Work

### A. Notations

We denote a vector $\mathbf{x}$ through a bold lower-case letter and a matrix $\mathbf{X}$ through a bold capital letter. An entry $x_n$ in a vector (or $X_{m,n}$ of a matrix) is denoted by a letter with a subscript that stands for its index. $\|\mathbf{x}\|_2$ represents the $l_2$ norm, $\|\mathbf{X}\|_*$ denotes the nuclear norm. We use $\cdot^T$ and $\cdot^H$ to denote the transpose and the conjugate transpose, respectively. The operators are denoted by calligraphic letters.

### B. Low-rank Property of the Hankel Matrix

The FID is denoted as $\tilde{\mathbf{x}} = [\tilde{x}_0, \ldots, \tilde{x}_{N-1}]^T \in \mathbb{C}^{N \times 1}$, where $N$ is the number of sampling points. By forwardly modeling the FID as a sum of $R$ damped exponential components [13], the $n^{th}$ entry of FID is expressed as:

$$\tilde{x}_n = \sum_{r=1}^{R} c_r e^{(i2\pi f_r - \tau_r)n\Delta t} = \sum_{r=1}^{R} c_r z_r^n, \quad (1)$$

where $i = \sqrt{-1}$, , and $z_r = e^{(i2\pi f_r - \tau_r)\Delta t}$, where $c_r$, $f_r$, $\tau_r$ and $\Delta t$ are the signal amplitude, central frequency, decay factor and sampling interval of the $r^{th}$ exponential function, i.e. spectral peak, respectively. $n \in [0, N-1]$ and $r \in [1, R]$.

$\tilde{\mathbf{x}}$ can be represented as the linear superposition of $R$ subspaces $\left\{ \mathbf{z}_r \in \mathbb{C}^{N \times 1} \middle| \mathbf{z}_r(n) = z_r^n = e^{(i2\pi f_r - \tau_r)n\Delta t} \right\}_{r=1,\ldots,R}$:

$$\tilde{\mathbf{x}} = \sum_{r=1}^{R} c_r \mathbf{z}_r, \quad (2)$$

By performing the Fourier transform on each $c_r \mathbf{z}_r$, one can obtain each physical spectral peak [2, 19]. In practice, $c_r \mathbf{z}_r$ is not known in advance and can be used to infer the chemical formula or protein structure.

Let $\mathcal{R}$ be a Hankel operator defined as $\mathcal{R}: \mathbb{C}^{N \times 1} \to \mathbb{C}^{P \times Q}$, $P = N - Q + 1$, which is a process of transforming a one-dimensional FID $\tilde{\mathbf{x}}$ into a Hankel matrix $\mathcal{R}\tilde{\mathbf{x}}$:

$$\mathcal{R}\tilde{\mathbf{x}} = \begin{bmatrix} \tilde{x}_0 & \tilde{x}_1 & \cdots & \tilde{x}_{Q-1} \\ \tilde{x}_1 & \tilde{x}_2 & \cdots & \tilde{x}_Q \\ \vdots & \vdots & \ddots & \vdots \\ \tilde{x}_{P-1} & \tilde{x}_P & \cdots & \tilde{x}_{N-1} \end{bmatrix}, \quad (3)$$

where the elements on each anti-diagonal of the Hankel matrix are equal. $Q$ is chosen as $(N+1)/2$ (or $N/2$) for an odd (or an even) $N$.

The low-rank property of the Hankel matrix $\mathcal{R}\tilde{\mathbf{x}}$ can be further understood through its SVD [20] according to

$$\mathcal{R}\tilde{\mathbf{x}} = \mathbf{U}\boldsymbol{\Sigma}\mathbf{V}^H, \quad (4)$$

where $\mathbf{U} \in \mathbb{C}^{P \times P}$ and $\mathbf{V} \in \mathbb{C}^{Q \times Q}$ are the left and right singular value matrices of $\mathcal{R}\tilde{\mathbf{x}}$, respectively, and they are both orthogonal matrices. $\boldsymbol{\Sigma} \in \mathbb{C}^{P \times Q}$ is a diagonal matrix, whose diagonal values arranged in the descending order are the singular values of the Hankel matrix $\mathcal{R}\tilde{\mathbf{x}}$.

The Hankel matrix can also be represented as a $R$-term linear superposition of the rank-1 matrix $\left\{ \tilde{\mathbf{X}}_r \in \mathbb{C}^{P \times Q} \middle| \tilde{\mathbf{X}}_r = \mathbf{u}_r \mathbf{v}_r^H \right\}_{r=1,\ldots,R}$ [2, 19, 21]:

$$\mathcal{R}\tilde{\mathbf{x}} = \sum_{r=1}^{R} \sigma_r \tilde{\mathbf{X}}_r = \sum_{r=1}^{R} \sigma_r \mathbf{u}_r \mathbf{v}_r^H, \quad (5)$$

where $\mathbf{u}_r$ and $\mathbf{v}_r$ are the $r^{th}$ column of $\mathbf{U}$ and $\mathbf{V}$, respectively; $\sigma_r$ represents the $r^{th}$ singular value of $\mathcal{R}\tilde{\mathbf{x}}$.

Now, we define an adjoint operator $\mathcal{R}^*: \mathbb{C}^{P \times Q} \to \mathbb{C}^{N \times 1}$ for $\mathcal{R}$, which transforms a matrix back to a vector by averaging the sum of elements on the anti-diagonal direction. $\mathcal{R}^* \tilde{\mathbf{X}}_r$ is called a virtual vector, and performing the Fourier transform on it yields a virtual spectral peak [2, 19]. $\sigma_r$ is corresponding to the virtual peak intensity [2, 11, 19], and a linear superposition of multiple virtual peaks can form the original spectral signal.

The rank of the Hankel matrix $\mathcal{R}\tilde{\mathbf{x}}$, i.e., the number of its non-zero singular values, is equal to the number of exponentials (or spectral peaks) $R$. Moreover, it commonly assumes that $R \ll N$, meaning that the number of spectral peaks is much smaller than the matrix size [2, 11-13]. Thus, the Hankel matrix $\mathcal{R}\tilde{\mathbf{x}}$ is low-rank. This property is called the low-rankness of the FID [2, 11-13].

### C. CHORD

In practical applications, an FID is contaminated by noise and the observed signal is a noisy one as [22]

$$\mathbf{y} = \tilde{\mathbf{x}} + \mathbf{n}, \quad (6)$$

where $\mathbf{n} \in \mathbb{C}^{N \times 1}$ represents the noise whose real and imaginary parts follow a Gaussian distribution with a mean 0 and a variance $\sigma^2$. Noise will make the matrix rank of $\mathcal{R}\mathbf{y}$ be larger than that of $\mathcal{R}\mathbf{x}$, and thus the noisy FID is no longer low-rank.

CHORD [10] introduces the low-rank property and tries to find a denoised FID that can balance its low-rankness and data fidelity according to

$$\min_{\mathbf{x}} \|\mathcal{R}\mathbf{x}\|_* + \frac{\lambda}{2}\|\mathbf{y} - \mathbf{x}\|_2^2, \quad (7)$$

where $\|\mathbf{x}\|_*$ denotes the nuclear norm, i.e. the sum of the singular values, and $\lambda$ is a regularization parameter.

Although CHORD has explored the low-rank property well and shown superior denoising performance over Cadzow and rQRD [10], it does not directly tap the exponential function property. This limitation may result in sub-optimal performance. For example, previous evidences found that the SVD on a Hankel matrix decomposes the spectrum into a same number of virtual spectral peaks but each of them is not exactly the same as the physical peaks [19]. Preserving the physical peaks of each exponential, through the Vandermonde decomposition, can obviously reduce the reconstruction error [15, 18]. Inspired by these works, our aim is to directly explore matrix decomposition of exponential function, the Vandermonde structure, to improve the denoising performance.

## III. PROPOSED METHOD

The main idea of the proposed method is to add an explicit constraint to decompose the FID into a Vandermonde matrix and preserve exponentials in denoising.

The proposed method is named as CHORD-V, which is short for CHORD with Vandermonde constraint. It not only inherits the low-rankness of Hankel matrix, but also explicitly enforces the Vandermonde structure to fits for exponential signal forms under the minimal norm principle.

### A. Model of CHORD-V

According to Eq. (2), the FID signal can be expressed by the multiplication of a Vandermonde matrix and a vector [18]:

$$\tilde{\mathbf{x}} = \sum_{r=1}^{R} c_r \mathbf{z}_r = \tilde{\mathbf{Z}}\tilde{\mathbf{c}} = \begin{bmatrix} z_1^0 & z_2^0 & \cdots & z_R^0 \\ z_1^1 & z_2^1 & \cdots & z_R^1 \\ \vdots & \vdots & \ddots & \vdots \\ z_1^{N-1} & z_2^{N-1} & \cdots & z_R^{N-1} \end{bmatrix} \begin{bmatrix} c_1 \\ c_2 \\ \vdots \\ c_R \end{bmatrix} \quad (8)$$

where $\tilde{\mathbf{Z}} \in \mathbb{C}^{N \times R}$ is a Vandermonde matrix whose column is $\left\{ \mathbf{z}_r \in \mathbb{C}^{N \times 1} \middle| \mathbf{z}_r(n) = z_r^n = e^{(i2\pi f_r - \tau_r)n\Delta t}, r = 1,...,R \right\}$, and $\tilde{\mathbf{c}} \in \mathbb{C}^{R \times 1}$ denotes the amplitudes of physical peaks.

We introduce the constraint of Vandermonde decomposition $\mathbf{x} = \mathbf{Z}\mathbf{c}$ into Eq. (7) and propose a model as follows:

$$\min_{\mathbf{x},\mathbf{c},\mathbf{Z}} \|\mathcal{R}\mathbf{x}\|_* + \frac{\lambda}{2}\|\mathbf{y} - \mathbf{x}\|_2^2 + \frac{\gamma}{2}\|\mathbf{c}\|_2^2 \quad (9)$$

s.t. $\mathbf{x} = \mathbf{Z}\mathbf{c}$, $\mathbf{Z}$ is a Vandermonde matrix

where $\mathbf{Z} \in \mathbb{C}^{N \times \hat{R}}$ is a Vandermonde matrix, $\lambda$ and $\gamma$ are regularization parameters, and $\|\mathbf{c}\|_2^2$ is defined as the power of $l_2$ norm and represents an energy constraint.

This model can be further expressed as:

$$\min_{\mathbf{x},\mathbf{c},\mathbf{Z}} \|\mathcal{R}\mathbf{x}\|_* + \frac{\lambda}{2}\|\mathbf{y} - \mathbf{x}\|_2^2 + \frac{\gamma}{2}\|\mathbf{c}\|_2^2 + \frac{\mu}{2}\|\mathbf{x} - \mathbf{Z}\mathbf{c}\|_2^2 \quad (10)$$

s.t. $\mathbf{Z}$ is a Vandermonde matrix

### B. Algorithm of CHORD-V

To solve the model in Eq. (10), the Alternating Direction Method of Multipliers (ADMM) algorithm is used [23].

Here, we define $\mathbf{X} = \mathcal{R}\mathbf{x}$, and Eq. (10) is written as:

$$\min_{\mathbf{x},\mathbf{c},\mathbf{X},\mathbf{Z}} \|\mathbf{X}\|_* + \frac{\lambda}{2}\|\mathbf{y} - \mathbf{x}\|_2^2 + \frac{\gamma}{2}\|\mathbf{c}\|_2^2 + \frac{\mu}{2}\|\mathbf{x} - \mathbf{Z}\mathbf{c}\|_2^2 \quad (11)$$

s.t. $\mathbf{Z}$ is a Vandermonde matrix

The augmented Lagrangian form of Eq. (11) is:

$$L(\mathbf{x},\mathbf{c},\mathbf{X},\mathbf{Z},\mathbf{D}) = \|\mathbf{X}\|_* + \frac{\lambda}{2}\|\mathbf{y} - \mathbf{x}\|_2^2 + \langle \mathbf{D}, \mathcal{R}\mathbf{x} - \mathbf{X} \rangle$$
$$+ \frac{\beta}{2}\|\mathcal{R}\mathbf{x} - \mathbf{X}\|_F^2 + \frac{\gamma}{2}\|\mathbf{c}\|_2^2 + \frac{\mu}{2}\|\mathbf{x} - \mathbf{Z}\mathbf{c}\|_2^2 \quad (12)$$

which is simplified as:

$$L(\mathbf{x},\mathbf{c},\mathbf{X},\mathbf{Z},\mathbf{D}) = \|\mathbf{X}\|_* + \frac{\lambda}{2}\|\mathbf{y} - \mathbf{x}\|_2^2$$
$$+ \frac{\beta}{2}\left\|\mathcal{R}\mathbf{x} - \mathbf{X} + \frac{\mathbf{D}}{\beta}\right\|_F^2 + \frac{\gamma}{2}\|\mathbf{c}\|_2^2 + \frac{\mu}{2}\|\mathbf{x} - \mathbf{Z}\mathbf{c}\|_2^2 \quad (13)$$

Then, the augmented Lagrangian form of Eq. (13) is:

$$\min_{\mathbf{x},\mathbf{c},\mathbf{X},\mathbf{Z}} \max_{\mathbf{D}} \|\mathbf{X}\|_* + \frac{\lambda}{2}\|\mathbf{y} - \mathbf{x}\|_2^2 + \frac{\beta}{2}\left\|\mathcal{R}\mathbf{x} - \mathbf{X} + \frac{\mathbf{D}}{\beta}\right\|_F^2$$
$$+ \frac{\gamma}{2}\|\mathbf{c}\|_2^2 + \frac{\mu}{2}\|\mathbf{x} - \mathbf{Z}\mathbf{c}\|_2^2 \quad (14)$$

s.t. $\mathbf{Z}$ is a Vandermonde matrix

Eq. (14) can be iteratively solved through multiple sub-problems:

1) $\mathbf{x}_{k+1}$ can be obtained by solving Eq. (15):

$$\min_{\mathbf{x}} \frac{\lambda}{2}\|\mathbf{y} - \mathbf{x}\|_2^2 + \frac{\beta}{2}\left\|\mathcal{R}\mathbf{x} - \mathbf{X} + \frac{\mathbf{D}}{\beta}\right\|_F^2 \quad (15)$$

Taking the derivative of Eq. (15) yields the solution:

$$\mathbf{x}_{k+1} = \left(\lambda \mathbf{I} + \mu \mathbf{I} + \beta \mathcal{R}^*\mathcal{R}\right)^{-1}\left(\lambda \mathbf{y} + \mu \mathbf{Z}_k \mathbf{c}_k + \mathcal{R}^*(\beta \mathbf{X}_k - \mathbf{D}_k)\right) \quad (16)$$

2) $\mathbf{Z}_{k+1}$ can be obtained by solving Eq. (17):

$$\min_{\mathbf{Z}} \frac{\mu}{2}\|\mathbf{x} - \mathbf{Z}\mathbf{c}\|_2^2, \quad (17)$$

The solution $\mathbf{Z}_{k+1}$ can be obtained from the SVD of the Hankel matrix while maintaining its shift-invariant structure as:

$$\mathbf{Z}_{k+1} = \mathcal{Z}(\mathcal{L}(\mathcal{T}_{\hat{R}}(\mathcal{R}\mathbf{x}_{k+1}))), \quad (18)$$

where $\mathcal{T}_{\hat{R}}(\mathcal{R}\mathbf{x}_{k+1})$ denotes truncating $\mathcal{R}\mathbf{x}_{k+1}$ to a rank $\hat{R}$ matrix, $\mathcal{L}(\cdot)$ is an operator that takes the left singular matrix, and $\mathcal{Z}$ is an operator to estimate a Vandermonde matrix from the left singular matrix. The detailed process of deriving and computing Eq. (18) is provided in appendix.

3) $\mathbf{c}_{k+1}$ is obtained through

$$\min_{\mathbf{c}} \frac{\gamma}{2}\|\mathbf{c}\|_2^2 + \frac{\mu}{2}\|\mathbf{x} - \mathbf{Zc}\|_2^2, \quad (19)$$

whose solution is:

$$\mathbf{c}_{k+1} = \left(\mu(\mathbf{Z}_{k+1})^H \mathbf{Z}_{k+1} + \gamma \mathbf{I}\right)^{-1}(\mu \mathbf{Z}_{k+1})^H \mathbf{x}_{k+1}. \quad (20)$$

4) $\mathbf{X}_{k+1}$ is obtained through

$$\min_{\mathbf{X}} \|\mathbf{X}\|_* + \frac{\beta}{2}\left\|\mathcal{R}\mathbf{x} - \mathbf{X} + \frac{\mathbf{D}}{\beta}\right\|_F^2, \quad (21)$$

whose solution is:

$$\mathbf{X}_{k+1} = \mathcal{S}_{1/\beta}\left(\mathcal{R}\mathbf{x}_{k+1} + \frac{\mathbf{D}_k}{\beta}\right), \quad (22)$$

where $\mathcal{S}(\cdot)$ denotes the singular value soft-thresholding operation defined as $\mathcal{S}_a(\mathbf{A}) = \mathbf{U}diag\{(\sigma_j - a)_+\}\mathbf{V}^H$.

5) $\mathbf{D}_{k+1}$ is updated according to

$$\mathbf{D}_{k+1} = \mathbf{D}_k + \tau(\mathcal{R}\mathbf{x}_{k+1} - \mathbf{X}_{k+1}), \quad (23)$$

where $\tau$ denotes step size.

The whole algorithm is summarized in Table 1, and it terminates if the consecutive difference of the solution, i.e. $\Delta\mathbf{x} = \|\mathbf{x}_{k+1} - \mathbf{x}_k\|_2 / \|\mathbf{x}_{k+1}\|_2$, is smaller than a tolerance $\eta$. This algorithm is observed empirically converged since $\Delta\mathbf{x}$ is gradually decreased in iterations (Fig. 1).

## IV. EXPERIMENTS

In this section, the proposed CHORD-V is compared with Cadzow, rQRd, and CHORD. Their denoising performances were evaluated on both synthetic and realistic NMR data. All computations were performed using MATLAB on a CentOS 7 compute server equipped with two Intel(R) Xeon(R) Silver 4210 CPUs (2.2 GHz, 10 cores in total), 256 GB of RAM, and eight Nvidia Tesla T4 GPUs.

To evaluate the denoising performance, the Normalized Root-Mean-Square Error (NRMSE) [24] is adopted as:

$$NRMSE = \|\hat{\mathbf{x}} - \tilde{\mathbf{x}}\|_2 / \|\tilde{\mathbf{x}}\|_2, \quad (24)$$

where $\tilde{\mathbf{x}}$ and $\hat{\mathbf{x}}$ are the noise-free and denoised signal, respectively.

**Table 1.** Algorithm of CHORD-V

**Input**: $\mathbf{y}$, $\lambda$, $\mu$; $\tau = \beta = 1$, the tolerance of solution in iterations $\eta = 10^{-3}$, the max iteration $K = 200$, the prior number of spectral peaks $\hat{R}$.

**Initialization**: $\mathbf{D}_0 = \mathbf{1}$, $\mathbf{Z}_0 = \mathcal{Z}(\mathcal{L}(\mathcal{T}_{\hat{R}}(\mathcal{R}\mathbf{y})))$, $\mathbf{X}_0 = \mathbf{0}$, $\mathbf{c}_0 = \left((\mathbf{Z}_{k+1})^H \mathbf{Z}_{k+1}\right)^{-1}(\mathbf{Z}_{k+1})^H \mathbf{y}$, $\mathbf{x}_0 = \mathbf{y}$, $\Delta\mathbf{x} = 10^5$, $k = 1$.

  **While** $\Delta\mathbf{x} > \eta$ and $k \leq K$, **do**:
  1) Update $\mathbf{x}_{k+1}$ according to Eq. (16);
  2) Update $\mathbf{Z}_{k+1}$ according to Eq. (18);
  3) Update $\mathbf{c}_{k+1}$ according to Eq. (20);
  4) Update $\mathbf{x}_{k+1} = \mathbf{Z}_{k+1}\mathbf{c}_{k+1}$;
  5) Update $\mathbf{X}_{k+1}$ according to Eq. (22);
  6) Update $\mathbf{D}_{k+1}$ according to Eq. (23);
  7) Compute $\Delta\mathbf{x} = \|\mathbf{x}_{k+1} - \mathbf{x}_k\|_2 / \|\mathbf{x}_{k+1}\|_2$;
  8) Update $k \leftarrow k+1$.
**End**
**Output**: $\mathbf{x}$

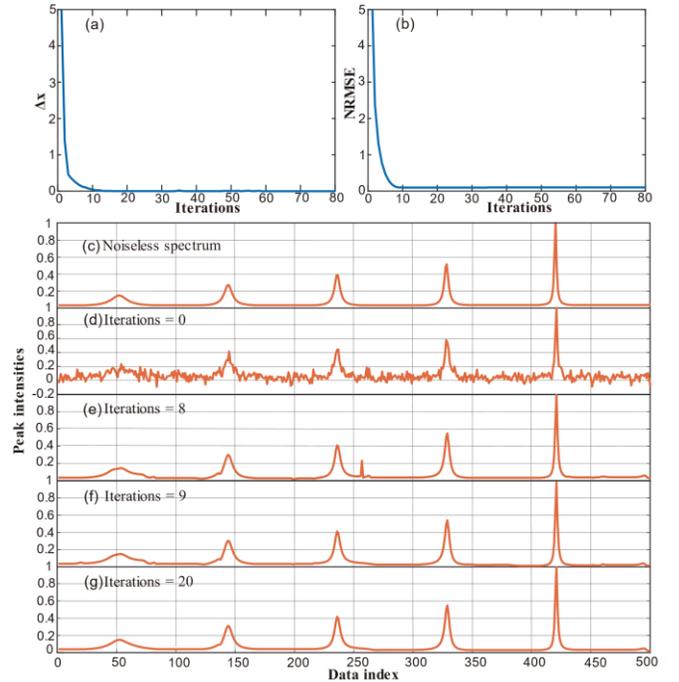

**Fig. 1.** Empirical convergence analysis of CHORD-V. (a) and (b) are the $\Delta\mathbf{x}$ and NRMSE in iterations, respectively. (c) is the noise-free spectrum, (d)-(g) are the denoising results of CHORD-V at iteration 0, 8, 9 and 20, respectively. Note: The noise standard deviation is 0.03, and the true number of peaks is 5 and we set the prior number $\hat{R}$=20 in simulations.

### A. Denoising Synthetic Data

The synthetic noise-free FID was generated by the superposition of multiple exponentials according to Eq. (1).

Then, Gaussian noise with a noise standard deviation σ was added to the FID to simulate a noisy FID.

The proposed CHORD-V obtains minimal errors than other methods under all noise levels (Fig. 2). When the noise is low (σ=0.02 in Fig. 2(a)), the denoising performances of Cadzow and CHORD-V are the best. As the noise increases, Cadzow tends to lose low-intensity peaks and introduce artifacts (Fig. 2(b)). rQRd is prone to residual noise. CHORD encounters an uneven baseline and some peak distortions. Although a slight peak distortion is observed in CHORD-V when the noise is high, its denoised spectra are closest to the noiseless spectrum under all noise levels.

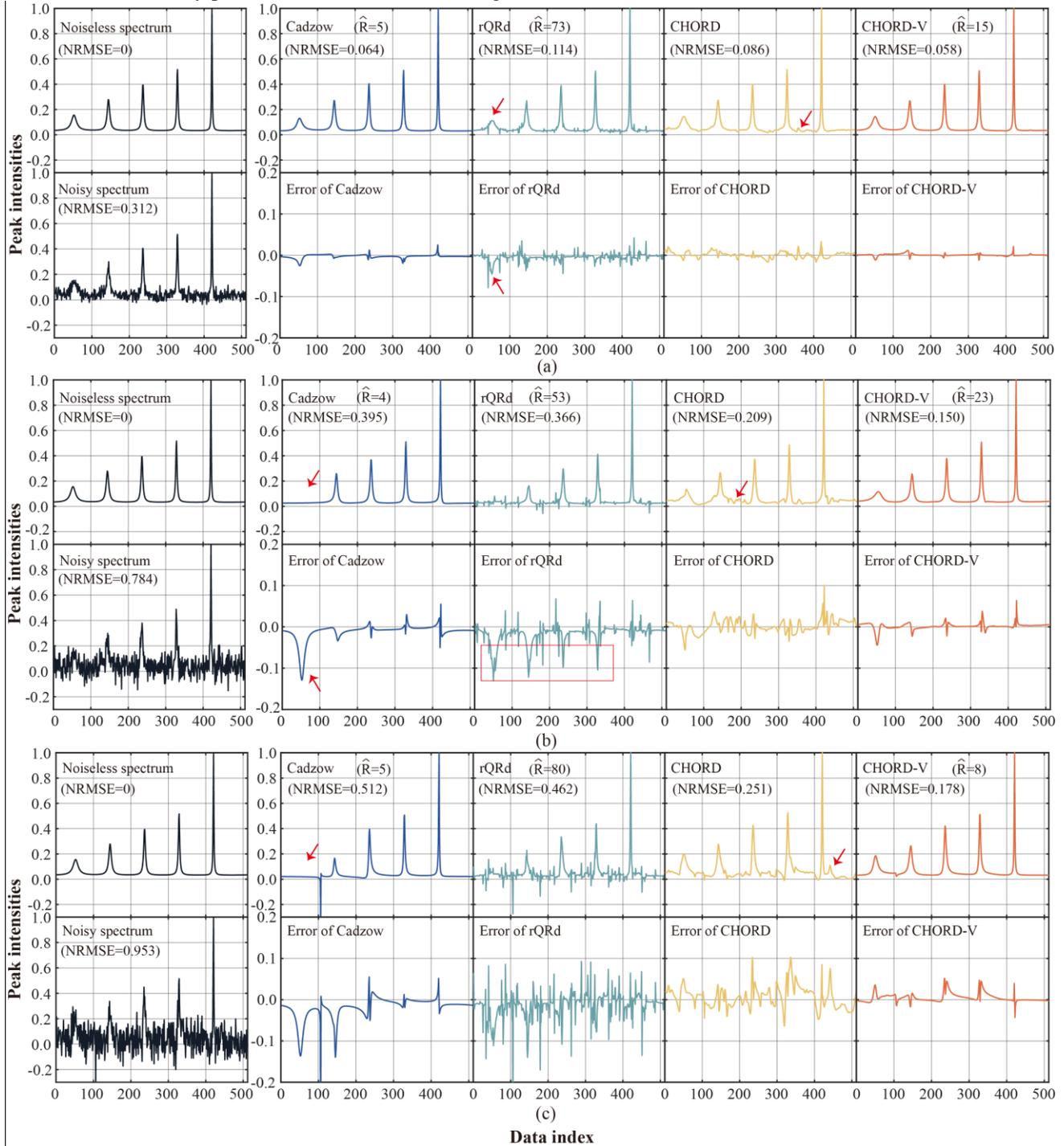

**Fig. 2.** Denoising results of synthetic data with noise standard deviations of 0.02 (a), 0.05(b) and 0.06(c), respectively. The blue, green, yellow, and red lines are denoised results by Cadzow, rQRd, CHORD and CHORD-V, respectively. Note: The prior numbers of spectral peaks $\hat{R}$ for Cadzow, rQRd and CHORD-V are determined by their respective minimal NRMSEs.

Fig.3 shows how the prior number of spectral peaks $\hat{R}$ affect denoising results. When $\hat{R}$ is equal to the ground-truth number of peaks, both Cadzow and rQRd will lose low-intensity peaks (Figs. 3(b) and (c)). Even with an optimal $\hat{R}$ in terms of minimal NRMSE, these two methods still introduce peak distortion or noise (Figs. 3(f) and (g)). CHORD does not require setting this parameter, achieving much lower NRMSE than Cadzow and rQRd, but compromises peak shapes (Fig. 3(i)). The proposed CHORD-V is more robust to $\hat{R}$ and can always obtain best denoising results (Figs. 3(d) and (h)). Even when $\hat{R}$ is 10 times of the true number, better denoised spectrum is achieved by CHORD-V than CHORD, particularly on preserving the lowest-intensity peak (Fig. 3(l)).

We further measure the denoising error NRMSE versus $\hat{R}$ in Fig. 4. Cadzow only reaches the lowest error when $\hat{R}$ is around 1~2 times of the true number of peaks and other settings will greatly increase the error. rQRd decreases the error by increasing $\hat{R}$. Both CHORD and CHORD-V significantly reduce the error, while the latter achieves relatively lower error when the noise standard deviation is high (0.05 and 0.06). CHORD-V is robust to $\hat{R}$, which only requires $\hat{R}$ be larger than the ground-truth number of peaks.

The robustness to noise levels is compared in Fig. 5. Under all compared noise levels, the average error of CHORD-V is always the lowest. When the noise standard deviation is between 0.01 and 0.06, the average denoising error of CHORD-V is reduced by 4%~65%, 37%~61% and 29%~36% than Cadzow, rQRd and CHORD, respectively. Besides, the error standard deviation of CHORD-V is 20%~75% of Cadzow, 12%~44% of rQRd and is slightly higher than CHORD by 0%~36%. Thus, CHORD-V maintains good robustness to the random noise, but the deviation is slightly larger than CHORD.

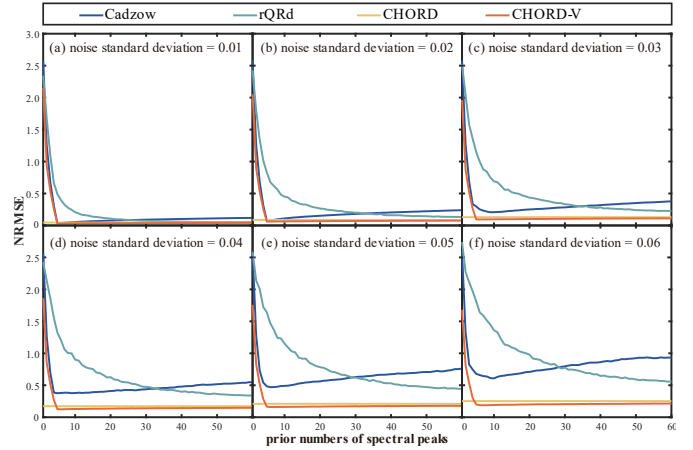

**Fig. 4.** NRMSE of the synthetic data with different prior numbers of spectral peaks $\hat{R}$. (a) - (f) are results when the noise standard deviation is from 0.01 to 0.06, respectively. Note: CHORD does not require setting $\hat{R}$. The curve represents the average NRMSE under 100 repeated simulations at the same noise level. The ground-truth number of peaks is 5.

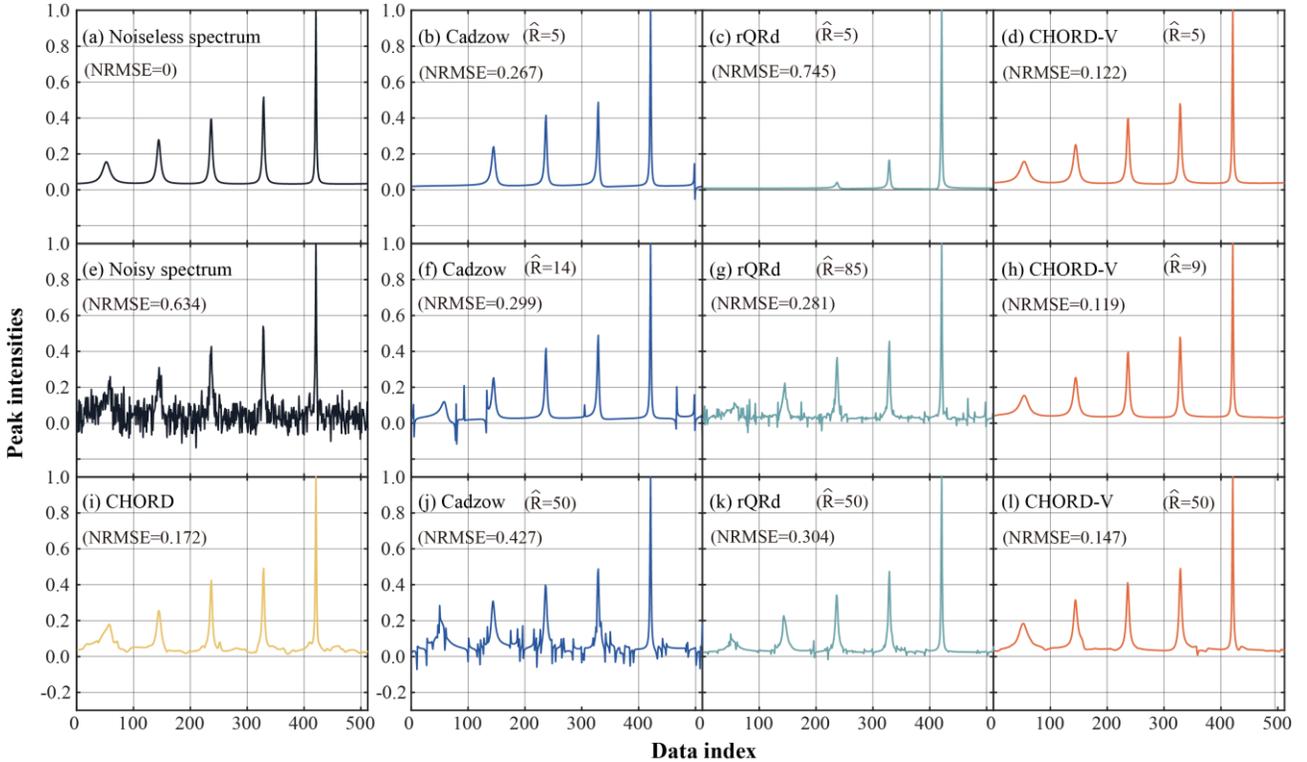

**Fig. 3.** Denoised synthetic spectra versus different prior numbers $\hat{R}$ of spectral peaks. (a) and (e) are the spectra without noise and with noise (the standard deviation is 0.04), respectively; (i) is the spectrum denoised by CHORD; (b) - (l) are spectra denoised by Cadzow, rQRd, and CHORD-V, respectively.

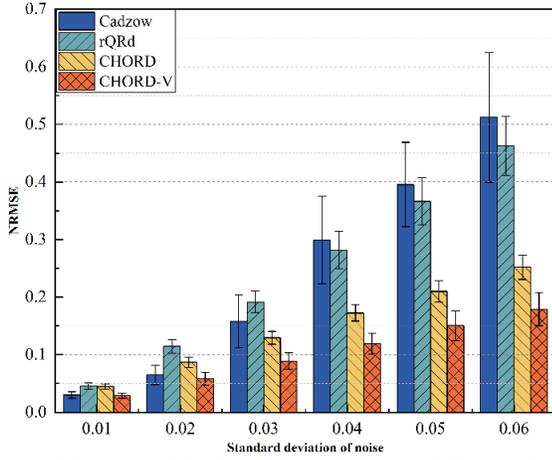

**Fig. 5.** Denoising errors of the simulated 5-peak signal at different noise levels. Note: The height of the bars and error bars represent the average and standard deviations of NRMSE under 100 repeated experiments, respectively. The prior numbers of spectral peaks used for all methods (except CHORD) are determined by their own minimum NRMSEs.

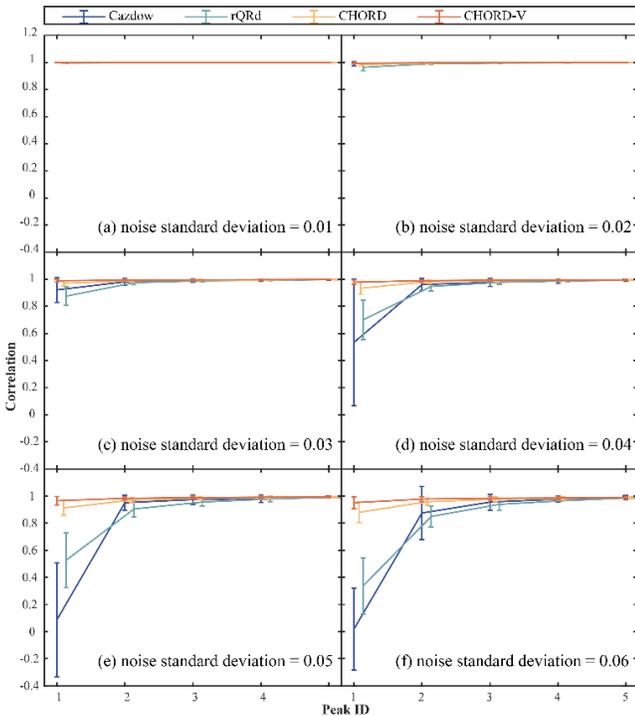

**Fig. 6.** Intensity correlations of each spectral peak between the noiseless and denoised spectrum for the synthetic spectrum with 5peaks. (a) - (f) are the results when the noise standard deviation is 0.01~0.06, respectively. Note: Peak ID corresponds to the five spectral peaks from left to right. The height of the line and error bars represent the average and standard deviations of correlations under 100 experiments, respectively. The prior numbers of spectral peaks for all methods (except CHORD) are determined by their own minimum NRMSEs.

Individual peak intensity correlations are measured in Fig. 6. All methods perform very well (correlations are close to 1) if the noise level is relatively high (Figs. 6(a) and (b)). As the noise increases, the correlation of medium-low intensity peaks (Peak ID = 1, 2) decreases, and the decrease of CHORD-V is much lower than those of other methods, with correlations closer to 1. When the noise standard deviation is high (0.04~0.06), for the low-intensity peak (Peak ID = 1), the average correlation of CHORD-V is increased by 83%~524%, 40%~182%, and 5%~8% than Cadzow, rQRd, and CHORD, respectively. For the medium-intensity peaks (Peak ID =2), the average correlation of CHORD-V is increased by 3%-12%, 5%-15%, and 1%-2%, respectively. Thus, both Cadzow and rQRd tend to distort low and medium intensity peaks, which are alleviated well by CHORD. The proposed CHORD-V outperforms other methods on maintaining peak intensities.

### B. NMR Spectroscopy Data

For real spectroscopy data, we acquired a one-dimensional $^1$H spectrum on a 500 MHz NMR spectrometer, using a single-pulse sequence with a pulse length of $8.3\mu s$ in a total acquisition time of 286.7s. The chemical sample contains creatine, choline, magnesium citrate and calcium citrate. To obtain the signal with a high SNR, we repeated the acquisitions 64 times and averaged them, obtaining a reference signal with little noise. Then, Gaussian noise with the standard deviations ranging from 0.005 to 0.050 was added to the reference signal, generating noisy signals with different levels of noise.

Fig. 7 shows that CHORD-V has advantage when the noise standard deviation is high (0.035 and 0.05). Under the noise standard deviation is 0.035 (or 0.05), CHORD-V reduces the denoising error by about 38% (or 44%), 23% (or 33%), 7% (or 10%) than Cadzow, rQRd and CHORD, respectively. Under smaller noise standard deviations (0.005 and 0.02), rQRd, CHORD and CHORD-V achieved very close performances.

Representative denoised NMR spectra are shown in Fig. 8. The real data contains overlapping peaks and low-intensity peaks. Cadzow encounters difficulties in recovering low-intensity peaks and may generate artifacts. rQRd loses low-intensity peaks and leaves a large amount of noise residue when the noise standard deviation is high. In all experiments, CHORD outperforms Cadzow and rQRd on removing noise and preserving peaks. However, CHORD may introduce an uneven baseline or artifacts, which could be treated as suspicious peaks (the 4$^{th}$ column in Fig. 8). In comparison, CHORD-V better preserves low-intensity peaks and overlapping peaks while effectively removing noise (the last column in Fig. 8). Besides, when $\hat{R}$ is 1~7 times of the true number of peaks, the NRMSE of CHORD-V remains stable (Fig. 9), indicating its robustness to $\hat{R}$.

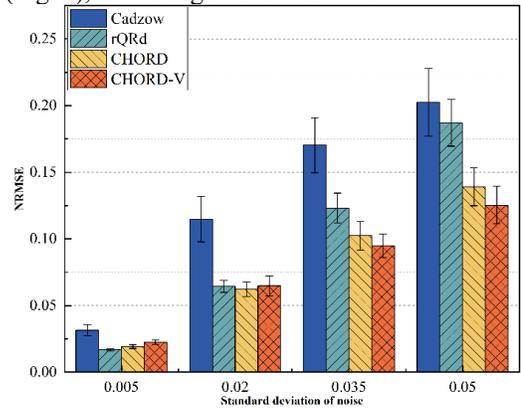

**Fig. 7.** Denoising errors of the real NMR spectroscopy data at different noise levels. Note: The height of the bars and error bars represent the average and standard deviations of NRMSE with 100 repeated denoising experiments, respectively. The prior numbers of spectral peaks for Cadzow, rQRd and CHORD-V are determined by their own minimum NRMSEs.

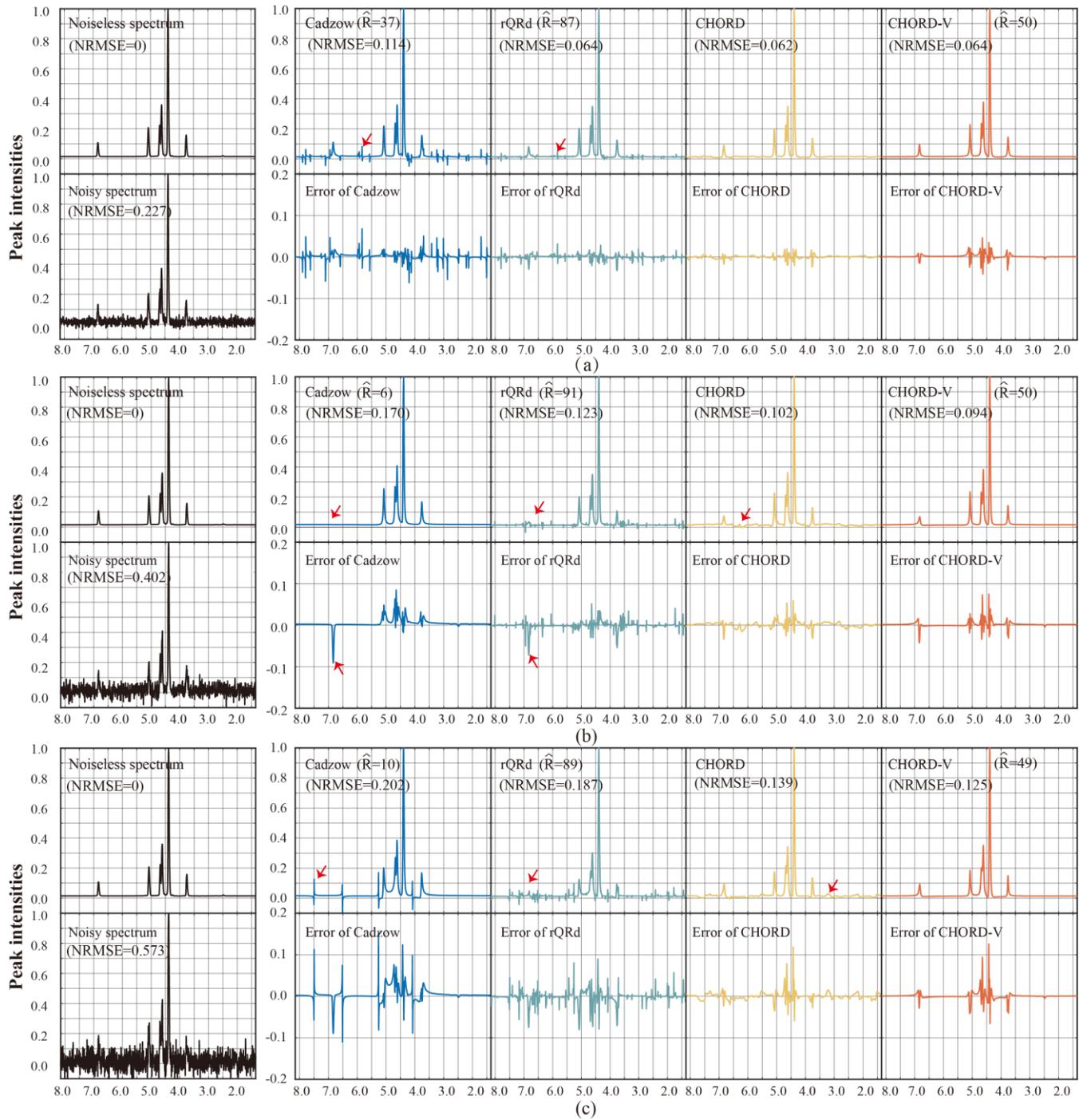

**Fig. 8.** Denoised NMR spectroscopy with the noise standard deviation of 0.020(a), 0.035(b) and 0.050(c), respectively. The blue, green, yellow, and red lines are denoised spectra by Cadzow, rQRd, CHORD and CHORD-V, respectively. Note: The prior numbers of spectral peaks for all methods (except CHORD) are determined by their own minimal NRMSEs.

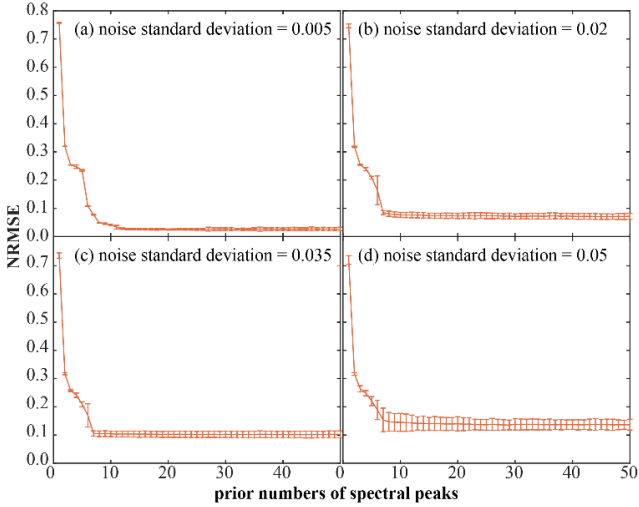

**Fig. 9.** Denoising errors of CHORD-V for the NMR spectroscopy data with different prior spectral peak numbers. (a)-(d) are denoising errors under the noise standard deviations of 0.005, 0.02, 0.035 and 0.05, respectively. Note: The height of the line and error bars represent the average and standard deviations of NRMSE from 100 repeated experiments, and the actual number of spectral peaks is 7.

## V. Conclusion

In this paper, an NMR spectra denoising method, CHORD-V, is proposed by modeling its time domain signal as superposition of exponentials and enforcing its Vandermonde structure in the low-rank reconstruction. Numerical experiments on synthetic and real NMR spectroscopy data show that this method is good at removing noise and preserving low-intensity peaks, particularly when the noise level is relatively high. Besides, the proposed method is robust to the number of prior spectral peaks in the algorithm.

## Appendix

$\mathbf{Z}_{k+1}$ can be obtained by solving Eq. (A1):

$$\min_{\mathbf{Z}} \frac{\mu}{2}\|\mathbf{x}-\mathbf{Z}\mathbf{c}\|_2^2 \quad s.t.\ \mathbf{Z} \text{ is a Vandermonde matrix} \quad (A1)$$

In the implementation, $\mathbf{Z}_{k+1}$ is obtained though SVD of the Hankel matrix and maintaining the shift-invariant structure of the matrix.

The Vandermonde decomposition [15, 18] of the Hankel matrix $\mathcal{R}\tilde{\mathbf{x}}$ is:

$$\mathcal{R}\tilde{\mathbf{x}} = \mathbf{E}\mathbf{\Lambda}\overline{\mathbf{E}}^T =$$

$$\begin{bmatrix} z_1^0 & z_2^0 & \cdots & z_R^0 \\ z_1^1 & z_2^1 & \cdots & z_R^1 \\ \vdots & \vdots & \ddots & \vdots \\ z_1^{P-1} & z_2^{P-1} & \cdots & z_R^{P-1} \end{bmatrix} \begin{bmatrix} c_1 & & 0 \\ & \ddots & \\ 0 & & c_R \end{bmatrix} \begin{bmatrix} z_1^0 & z_2^0 & \cdots & z_R^0 \\ z_1^1 & z_2^1 & \cdots & z_R^1 \\ \vdots & \vdots & \ddots & \vdots \\ z_1^{Q-1} & z_2^{Q-1} & \cdots & z_R^{Q-1} \end{bmatrix}^T$$

(A2)

where $\mathbf{E} \in \mathbb{C}^{P\times R}$ and $\overline{\mathbf{E}} \in \mathbb{C}^{Q\times R}$ are Vandermonde matrices, and $\mathbf{\Lambda} = diag\{c_r\} \in \mathbb{C}^{R\times R} (r=1,...,R)$ is a diagonal matrix.

This Vandermonde matrix has a shift-invariant structure [25]:

$$\mathbf{E}^f = \mathbf{E}^l diag(z_1, z_2, ..., z_R), \quad (A3)$$

where $(\cdot)^f$ and $(\cdot)^l$ denotes removing the first and last row of the matrix, respectively. Eq. (A3) can be expressed as:

$$(\mathbf{E}^l)^{-1}\mathbf{E}^f = diag(z_1, z_2, ..., z_R) \quad (A4)$$

where $(\cdot)^{-1}$ denotes the pseudoinverse and $diag(z_1, z_2, ..., z_R)$ is a diagonal matrix whose diagonal elements are eigenvalues.

Thus, once eigenvalues of the left-hand side are estimated, one can construct the subspaces $\{\mathbf{z}_r \in \mathbb{C}^{N\times 1}\}_{r=1,...,R}$, i.e. the exponential function, of FID $\tilde{\mathbf{x}}$ and a Vandermonde matrix

$$\tilde{\mathbf{Z}} = \begin{bmatrix} z_1^0 & z_2^0 & \cdots & z_R^0 \\ z_1^1 & z_2^1 & \cdots & z_R^1 \\ \vdots & \vdots & \ddots & \vdots \\ z_1^{N-1} & z_2^{N-1} & \cdots & z_R^{N-1} \end{bmatrix}.$$

Let $\mathcal{Z}$ denote an operator $\mathcal{Z}:\mathbb{C}^{P\times R} \to \mathbb{C}^{N\times R}$ to construct a Vandermonde matrix from a given matrix $\mathbf{E}$ as

$$\tilde{\mathbf{Z}} = \mathcal{Z}(\mathbf{E}) \quad (A5)$$

where $\mathcal{Z}(\mathbf{E})$ contains three steps, including calculating $\mathbf{A} = (\mathbf{E}^l)^{-1}\mathbf{E}^f$, estimating eigenvalues of matrix $\mathbf{A}$, and using eigenvalues to construct a Vandermonde matrix $\tilde{\mathbf{Z}}$.

After truncating the SVD decomposition of the Hankel matrix $\mathcal{R}\mathbf{x}$, we define a truncated matrix as:

$$\tilde{\mathbf{X}}_R = \mathcal{T}_R(\mathcal{R}\tilde{\mathbf{x}}) = \mathbf{U}_R \mathbf{\Sigma}_R \mathbf{V}_R^H \quad (A6)$$

where $\mathcal{T}_R(\mathcal{R}\tilde{\mathbf{x}})$ denotes truncating $\mathcal{R}\tilde{\mathbf{x}}$ to be rank $R$, $\mathbf{U}_R \in \mathbb{C}^{P\times R}$, $\mathbf{\Sigma}_R \in \mathbb{C}^{R\times R}$ and $\mathbf{V}_R \in \mathbb{C}^{Q\times R}$ denote extract the first $R$ columns of $\mathbf{U}$, $\mathbf{\Sigma}$ and $\mathbf{V}$, respectively.

Since Eq. (A2) and Eq. (A6) are truncating $\mathcal{R}\tilde{\mathbf{x}}$ to be rank $R$, $\mathbf{U}_R$ and $\mathbf{E}$ span the same column space, meaning that there exists an invertible matrix $\mathbf{G}$ to $\mathbf{U}_R = \mathbf{E}\mathbf{G}$. Thus,

$$\begin{cases} \mathbf{U}_R^f = \mathbf{E}^f \mathbf{G} \\ \mathbf{U}_R^l = \mathbf{E}^l \mathbf{G} \\ \mathbf{E}^f = \mathbf{E}^l diag(z_1, z_2, ..., z_R) \end{cases}, \quad (A7)$$

which means $(\mathbf{U}_R^l)^{-1}\mathbf{U}_R^f = \mathbf{G}^{-1}diag(z_1, z_2, ..., z_R)\mathbf{G}$. Thus, estimating the eigenvalues of $(\mathbf{U}_R^l)^{-1}\mathbf{U}_R^f$ can obtain the subspace $z_r, r=1,...,R$ of the FID signal $\tilde{\mathbf{x}}$. Therefore,

$$\tilde{\mathbf{Z}} = \mathcal{Z}(\mathbf{U}_R). \quad (A8)$$

In summary, one has

$$\mathbf{Z}_{k+1} = \mathcal{Z}(\mathcal{L}(\mathcal{T}_{\hat{R}}(\mathcal{R}\mathbf{x}_{k+1}))), \quad (A9)$$

where $\mathcal{T}_{\hat{R}}$ is defined in Eq. (A6), $\mathcal{L}(\cdot)$ is an operator which extracts the left singular matrix of $\mathcal{T}_{\hat{R}}(\mathcal{R}\mathbf{x}_{k+1})$, and $\mathcal{Z}$ is defined in Eq. (A5).